\def\lsim{\lower.7ex\hbox{${\buildrel < \over \sim}$}}
\def\gsim{\lower.7ex\hbox{${\buildrel > \over \sim}$}}
\documentstyle[12pt,epsf]{article}
\begin{document}

\centerline {Very High Energy Gamma Rays from  PSR1706-44}

\bigskip

\centerline{T.Kifune$^{(1)}$, T.Tanimori$^{(2)}$, S.Ogio$^{(2)}$,
T.Tamura$^{(1)}$, H.Fujii$^{(3)}$, M.Fujimoto$^{(4)}$,}

\centerline{T.Hara$^{(5)}$, N.Hayashida$^{(1)}$, S.Kabe$^{(3)}$,
F.Kakimoto$^{(2)}$, Y.Matsubara$^{(6)}$, Y.Mizumoto$^{(7)}$,}

\centerline{Y.Muraki$^{(6)}$, T.Suda$^{(7), *}$, M.Teshima$^{(1)}$,
T.Tsukagoshi$^{(2)}$, Y.Watase$^{(3)}$, T.Yoshikoshi$^{(2)}$,}

\centerline{P.G.Edwards$^{(8)}$, J.R.Patterson$^{(8)}$, M.D.Roberts$^{(8)}$,
G.P.Rowell$^{(8)}$ and G.J.Thornton$^{(8)}$}

\bigskip

\noindent 1. Institute for Cosmic Ray Research, University of Tokyo, Tokyo 188,
Japan

\noindent 2. Department of Physics, Tokyo Institute of Technology, Tokyo 152,
Japan

\noindent 3. National Laboratory for High Energy Physics (KEK), Tsukuba 305,
Japan

\noindent 4. National Astronomical Observatory, Tokyo 181, Japan

\noindent 5. Yuge National College for Maritime Technology, Ehime 794-25, Japan

\noindent 6. Solar-Terrestrial Environment Laboratory, Nagoya University,
Nagoya 464, Japan

\noindent 7. Department of Physics, Kobe University, Hyogo 637, Japan

\noindent 8. Department of Physics and Mathematical Physics, University of
Adelaide, South Australia 5005, Australia

\medskip

\noindent * Deceased

\vskip5truemm
\baselineskip=0.8cm

\centerline{Abstract}

\medskip

We have obtained
evidence of gamma-ray emission above 1 TeV from  PSR1706-44,
using a ground-based telescope of the atmospheric \v Cerenkov imaging type
located near Woomera, South Australia.
This object,  a $\gamma$-ray source
discovered by the COS B satellite (2CG342-02),
was identified with the radio pulsar
through the discovery of a 102  ms pulsed signal
with the EGRET instrument
 of the Compton Gamma Ray Observatory.
The flux of the present observation above a threshold of 1 TeV is
$\bf \sim $ 1 $\cdot$ 10$^{-11}$ photons  cm$^{-2}$ s$^{-1}$,
which is two orders of magnitude smaller than
the extrapolation from GeV energies.
The analysis is not restricted to a search for emission modulated
with the 102 ms period, and the reported flux is for all
$\gamma$-rays from PSR1706-44, pulsed and unpulsed.
The energy output in the TeV region corresponds to
about 10$^{-3}$ of
the spin down energy loss rate of the neutron star.


\noindent {\it Subject headings}: gamma rays: observations -- pulsars:
 individual (PSR1706-44) -- supernova remnants

\bigskip

\eject

\noindent 1. Introduction

\medskip

To date, five young pulsars -- the Crab (Swanenburg et al. 1981;
Nolan et al. 1993),
Vela (Swanenburg et al. 1981), Geminga (Bertsch et al. 1992),
PSR1706-44 (Kniffen et al. 1992; Thompson et al. 1992)
and PSR1055-52 (Fierro et al. 1993) --
have been discovered to be bright
in the hundreds of MeV to GeV $\gamma$-ray energy region.
Inspection of these five reveals that the longer the pulsar period,
the greater the fraction of the total spin down energy loss
that is emitted in high energy $\gamma$-rays.
The Crab nebula, containing the pulsar with the shortest period of these five,
is so far the only established source in which $\gamma$-ray emission extends
to the very high energy region near 1 TeV (Weekes et al. 1989;
Vacanti et al. 1991)
and even higher (Baillon et al. 1993; Tanimori et al. 1994).
More observational data are needed to understand the processes
in these objects
which result in greater energy losses at very high energies than any other.

\medskip

The pulsar PSR1706-44 has been observed with the 3.8m diameter
\v Cerenkov imaging telescope of
the CANGAROO Collaboration (Patterson and Kifune 1992; Hara et al. 1993;
Gregory et al. 1990) near Woomera, South Australia.
This object was detecteded as a $\gamma$-ray source (2CG342-02)
with the COS B satellite (Swanenburg et al. 1981)
and then identified with the radio pulsar
through the discovery of 102  ms pulsed signal
with the EGRET instrument of the Compton Gamma Ray Observatory
(Kniffen et al. 1992; Thompson et al. 1992).
The slowing down of the pulsar period indicates that the pulsar has
 a characteristic age of 17,000 years.
 X-ray emission was also found with the ROSAT X-ray satellite
(Becker et al. 1992), and this may suggest
an X-ray synchrotron nebula associated with the pulsar PSR1706-44
similar to the case of the Crab.
McAdam, Osborne and Parkinson (1993) recently suggested
a possible association of the pulsar with a shell-type supernova remnant.
Although these features are somewhat different from  the Crab case,
a common mechanism
of rapid rotation of pulsar magnetosphere is likely to cause
VHE $\gamma$-ray emission also in PSR1706-44.
Cheng and Ding (1994) discussed the energy spectra
of  $\gamma$-ray pulsars
at 100 MeV to 10 GeV energies
to fit them with parameters based on
the `outer gap' model of the pulsar magnetosphere.
The best fit value they have found so far for the
$\gamma$-ray emission distance
from the neutron star
suggests that PSR1706-44 should emit double pulses and could be
a TeV $\gamma$-ray emitter.

\medskip

Evidence of TeV $\gamma$-ray emission was found
in data from July and August 1992 (Ogio et al. 1993)
and in order to confirm this result observations were undertaken
 in 1993 (Kifune et al. 1993a).
This paper presents the results from these two years of observations.

\bigskip

\noindent 2. Instrumentation and Analysis Method

\bigskip

Atmospheric \v Cerenkov telescopes detect optical \v Cerenkov photons
emitted as air showers
initiated by very high energy $\gamma$-rays and cosmic rays
travel through the upper atmosphere.
The total number of \v Cerenkov photons is proportional to the
$\gamma$-ray energy.
The 3.8 m, alt-azimuth mounted, telescope (Hara et al. 1993) has
a  $\sim$ 1 TeV threshold for detecting $\gamma$-rays, and
is  equipped with a multi-pixel camera
of 220 photomultiplier tubes,
each of which views a $0.^{\circ}12 \times 0.^{\circ}12$ area of the sky,
with a total camera field of view of about 3$^{\circ}$.
The camera measures the number of photo-electrons
and  the pulse arrival time in each tube.
The event trigger requires more than three
photomultiplier tubes to have a signal of $>$ 3 photo-electrons
and the total number of photo-electrons detected to be
larger than about 20.
The arrival time measurement in each photomultiplier tube
is  useful for distinguishing the background, randomly incident, light
in contrast to the nearly simultaneous \v Cerenkov photons.
The night sky background light is estimated to be about 0.05 photo electrons
per 10 ns per photomultiplier tube (Hara et al. 1993), and,  thus,
the background light which affects
the \v Cerenkov light image consisting  typically of
10 photomultiplier tubes is  $\sim 1$ photo-electron.
A bright star, of magnitude 3.3 ($\eta \, Scorpii$), is located
about 1.4$^{\circ}$ from PSR1706-44,
 within the field of view of the camera.
By monitoring the singles count rate in each phototube,
we were able to monitor the passage of this star around the field of view, and
 the pointing of the telescope was  calibrated.
The object PSR1706-44 was kept within 0.$^{\circ}$1
 of the centre of the field of view.

\medskip

Monte Carlo simulations indicate that the image shape of
\v Cerenkov light from a $\gamma$-ray shower can be approximated by
 an ellipse, which is elongated towards the source position.
A cosmic ray shower produces a broader, more irregularly shaped image.
The image  can be characterised by
parameters such as the
\lq width', \lq length',
\lq distance' (the distance between the centroid of the image
and the centre of the field of view)
and the orientation of the image, \lq$\alpha$'.
The parameter $\alpha$ is the angle between
the major axis of the elliptical image
and the line
joining the nominal source position (usually the centre of the field of view)
to the centroid of the observed image.
The definition of these parameters used in our analysis
is the same as those of Weekes et al. (1989).
The Whipple detections of very high energy $\gamma$-rays
from the Crab nebula (Vacanti et al. 1991; Lewis et al. 1993) and
the galaxy Markarian 421 (Punch et al. 1991)
have demonstrated that the $\gamma$-ray signal can be distinguished from
the cosmic ray background
using the image shape. They have shown
that $\gamma$-ray events from a point source  appear as a peak near the origin
on a flat background distribution
when event rate is plotted as a function of $\alpha$.
The size of this peak will be affected by the choice of the nominal source
position used in the determination of $\alpha$, so that the position
of a $\gamma$-ray point source
within the field of view can be determined to an accuracy
of $\sim$ 0.$^{\circ}$1.

The procedures of analysing \v Cerenkov images
 applied to telescopes with different characteristics
 can be examined by using the Crab pulsar/nebula
 as a standard candle of very high energy $\gamma$-rays.
The result of the Crab observations (Tanimori et al. 1994)
with the present telescope was found to be compatible
with the energy spectrum reported by the Whipple group (Lewis et al. 1993).

\bigskip

\noindent 3. Observed Data

\bigskip

The image parameters were  calculated
for the events that are located well inside,
but not at the centre of, the field of view,
selecting the events with \lq distance' smaller than 0.$^{\circ}$9
and greater than the \lq length'.
Those events with compact images
(narrower than energy dependent values typically of
 0.$^{\circ}$18 for `width' and 0.$^{\circ}$45 for `length')
were then selected to enrich any $\gamma$-ray component present.
These values for selecting data differ slightly from those
applied to the Crab data
(0.$^{\circ}$14 for `width' and 0.$^{\circ}$33 for `length'
(Tanimori et al. 1994)).
Larger discrimination values are used in the present analysis
as the image size is smaller at
the larger zenith angles
of the Crab observations (Kifune et al. 1993b).
The discrimination values employed here result in
a similar fraction of $\gamma$-rays
 passing the selection cut as for the Crab analysis (Tanimori et al. 1994).
The average image size of the Crab observations was about 3/4 of
that at the zenith, consistent with
a Monte Carlo simulation for inclined showers (Tsukagoshi 1994).
Comparison with simulations as well as the dependence
of the chosen selection criteria on the total yield of photons
will be described in a subsequent paper.

The number of analysed events is plotted  in Figure 1 as a function of
the parameter \lq$\alpha$'.
Figure 1a is the plot for all the data sets combined, and
Figure 1b, $1c$ and $1d$ are from the 1992 data set, July 1993 data set
and August 1993 data set, respectively.
In order to monitor the cosmic ray background in `on-source'
data, `off-source' observation runs were also done.
In the on-source run, PSR1706-44 was tracked at the centre of
the field of view, whereas for off-source runs a point
 at the same declination as PSR1706-44
 but offset in right ascension was tracked.
The solid line in the figure indicates the on-source data, and
the dotted line shows the off-source data.
A peak of events is seen  at $\alpha \sim$ 0$^{\circ}$
for the on-source data but not for the off-source data.
In 1992, 18 hours of off-source data
and 42 hours of on-source data were recorded
and the results shown in Figure 1b.
For the 1993 data set,
the observation time is 42 hours for both on- and off-source runs.
The event rates of on- and off-source observations were
equal to each other
at $\alpha \, >$ 30$^{\circ}$.
The peak at $\alpha \sim$ 0$^{\circ}$ is
 thus due to an excess of gamma ray events above
 the background of isotropic cosmic rays.
The excess  at $\alpha \sim$ 0$^{\circ}$ is present
also when no selection cut  was made on `width', and
becomes more prominent as narrower widths are selected,
as expected if a $\gamma$-ray signal is present.
The width of the  observed peak is consistent with that estimated
from Monte Carlo simulations for $\gamma$-ray emission from a
point source.

\medskip

An analysis was performed in which the `source' position
was deliberately shifted (in software) from  PSR1706-44.
This procedure leads to a change in calculated values of `$\alpha$'.
We then studied
how the peak strength at $\alpha \sim$ 0$^{\circ}$ for these `false sources'
varied
as a function of the shift.
The contour map of the significance of the peak
(calculated from the number of events in $\alpha \, < \, 10^{\circ}$
compared with $\alpha \, > \, 30^{\circ}$)
is shown in Figure 2 as a function of the position around PSR1706-44.
Since the telescope is alt-azimuth mounted,
the field of view rotates around its center with time.
The `de-rotation' to correct for this effect was made to the data in software.
The highest significance is obtained when the source location is set
at the centre of the field of view, the true position of PSR1706-44.

The number of data samples in each map in Figure 2 is 1681.
We would therefore expect a few $\pm$3$\sigma$ statistical fluctuations.
However, deviations of about 4$\sigma$ are evident near the outer regions
of the map.
In the analysis to obtain Figure 2,
the requirement that `distance' be smaller than 0.$^{\circ}$9 was applied
to the `distance' of the centroid of image from each `false source' position
as well as the usual `distance' from the center of field of view.
This procedure resulted in unequal areas in the field of view
which contribute to the data for different assumed source positions,
and led to poorer statistics for the `false source'
positions near the edge of the field of view.
A strict comparison between `source' and `false source' should be made
to similarly sized data-bases (as proposed in Fomin et al. 1994).
In the present case, this condition is met only at the center of
the field of view.
However, the inequality of the angular area is not large for
the positions near the center, and `false sources' in this central region
do not produce fluctuations larger than expected from statistics.
The  significance
obtained for the true position of PSR1706-44
is the only one that exceeds statistical expectations near the center.
The existence of $4\sigma$ fluctuations near the edges of the map
suggests a limitation to this contour map method
when applied to `false source' locations considerably removed from the
center of the field of view.

The selection of `distance' less than $0.^{\circ}$9 is
found effective in suppressing any effects of the tendency for
images located near the edge of field of view
to appear elongated along
the outer boundary of the camera,
{\it i.e.} having values of $\alpha$ near $90^{\circ}$.
Enrichment of $\alpha \, > \, 30^{\circ}$ events as a consequence
of this effect would result in
the number of events at $\alpha \sim 0^{\circ}$ becoming
smaller than the rate estimated from $\alpha$ of larger values.

We examined the data for the presence of any systematic effect.
The events that contribute to the peak
are distributed over the full range of
$\gamma$-ray energy as estimated
from the total yield of \v Cerenkov photons.
Thus, the  peak at $\alpha \sim$ 0$^{\circ}$ is not
caused by a spurious effect near the limit of
detection sensitivity.
The plot of the centroid positions in the image plane
shows a uniform distribution
indicating no correlation with any specific position,
 such as that of $\eta \, Scorpii$.
This star is offset from PSR1706-44 by 0.$^{\circ}$64 in right ascension
and by 1.$^{\circ}$25 in declination:
a location beyond the upper right edge
in the map of Figure 2.
The separation of this star from PSR1706-44 is greater than the
$0.^{\circ}$9 discrimination value on `distance',
and, thus, it is unlikely that
the analysed images are affected by $\eta \, Scorpii$.

\medskip

The event rate of the cosmic ray background has a flat distribution
as a function of the parameter $\alpha$.
By estimating the background
 from the rate in the range  $\alpha$ = 30$^{\circ} \sim$ 90$^{\circ}$,
 the excess counts within the bins with $\alpha <$ 10$^{\circ}$ in Fig.$\, 1a$
 has a significance of 12 $\sigma$.
The number of events in the peak
 is  constant  within  statistical fluctuations
when the data are divided into shorter sub-sets.
The $\gamma$-ray intensity is, therefore, consistent with
steady emission during the observation period.

\medskip

The effective detection area  is a product of
the area of the \v Cerenkov light pool
and the detection efficiency of the telescope.
The detection efficiency is
a function of $\gamma$-ray energy and
(the location of telescope in)
the lateral spread  of a \v Cerenkov light pool,
 and also depends on the trigger conditions
and on the selection criteria of events for analysis.
Results of a Monte Carlo simulation study under way
to estimate the detection efficiency for these observations, suggest an
effective area of approximately 5$\times$ 10$^8$ cm$^2$
at a $\gamma$-ray energy of 1 TeV.
Adopting this area, the flux of $\gamma$-ray events
is $7 \cdot \eta ^{-1} \times 10^{-12}$ cm$^{-2}$ s$^{-1}$,
where $\eta$ is the efficiency for
 $\gamma$-rays passing through the data reduction procedures.
The number of excess counts which constitute the peak
at $\alpha \sim 0^{\circ}$
was compared between the two cases with and without the selection cuts
on `width' and `length'.
Ninety per cent of the peak events  survive after the cuts,
 similar to the case of the Crab data (Tanimori et al. 1994).
By putting $\eta$ equal to  0.9,
 we obtain a $\gamma$-ray flux  of
$ 8 \times 10^{-12}$ cm$^{-2}$ s$^{-1}$,
integrated above the threshold energy of 1 TeV.
We estimate
that the systematic error in the flux
and $\gamma$-ray energy may be as large as $\sim 30$ \%
from  uncertainties in the adopted
values of effective area and efficiency $\eta$.
Searches for pulsed emission correlated with the 102 ms period
(in preparation for publication) indicate that
it is unlikely that a major portion of the detected flux is
pulsed.

\bigskip

\noindent 4. Discussions

\bigskip

The present  yield of $\gamma$-rays corresponds to $\sim 1 \cdot 10^{-11}$
erg cm$^{-2}$ s$^{-1}$ at 1 TeV.
For a distance of 1.5 kpc to PSR1706-44,
 the TeV luminosity is
 $\sim 3 \cdot 10^{33}$ erg s$^{-1}$
assuming isotropic emission.
This is $10^{-3}$  of the loss rate
of the total spin down energy of
 $3.4 \cdot 10^{36}$ erg  s$^{-1}$ of the neutron star.
The TeV luminosity is smaller by an order of magnitude
than the luminosity of
 $2.6 \cdot 10^{34}$ erg s$^{-1}$ in the 100 MeV -- 10 GeV range,
but greater than the X-ray luminosity
of $1 \cdot 10^{32}$ erg  s$^{-1}$ in the 0.1 -- 2.4 keV  range
 detected with the ROSAT.
The energy spectrum of the pulsed signal reported in Thompson et al. 1992
is very flat up to several GeV
with a differential spectral index of $-$1.7.
When extrapolated to   1  TeV
 the integral flux is as high as
$1.6 \times 10^{-9}$ photons  cm$^{-2}$ s$^{-1}$
(time-averaged over the period).
 The flux at 1 TeV from the present observation is
smaller by two orders of magnitude than
 this extrapolated value.

\medskip

We note that
Nel et al. (1993) have  reported an upper limit of
$ 5.8 \times 10^{-12}$ cm$^{-2}$ s$^{-1}$
for $> \, 2.6$ TeV on the $\gamma$-ray flux
that is modulated with the pulsar period.
However, our analysis is not restricted to a search for pulsed emission,
and is sensitive to any $\gamma$-rays from PSR1706-44 --
pulsed and unpulsed.

\medskip

It has been reported by McAdam, Osborne and Parkinson (1993)
 that PSR1706-44 is located near
the edge of a shell-shaped supernova remnant
of about a half degree diameter.
If the production mechanism of TeV $\gamma$-rays is similar to
that of the Crab,
then pulsar PSR1706-44 may also have a
 synchrotron X-ray nebula like the Crab, and
 the ROSAT result may be an indication of this.
Although the peak of the $\alpha$-distribution is
consistent with a point source,
the angular resolution of the telescope,
$\sim 0.1^{\circ}$, corresponding to
 a size  $\lsim$ 3 pc at a distance of 1.5 kpc,
does not preclude such extended emission.
Multi-wavelength studies of this object will be needed in order to investigate
the production mechanism for the very high energy $\gamma$-rays
reported here.

\medskip

This work is supported by a Grant-in-Aid in Scientific Research from the Japan
Ministry of Education, Science and Culture,
and also by the Australian Research Council and International Science and
Technology Program. PGE acknowledges the receipt of a QEII Fellowship, and GJT
an ARC Fellowship.

\medskip
\vfill
\eject

\noindent References
\vspace{12pt}

Baillon, P., et al. 1993,
 {\it Proc. 23rd Int. Cosmic Ray Conf. (Calgary)}, {\bf 1}, 27

Becker, W. et al., 1992, {\it IAU Circular}, {\bf 5554}

Bertsch, D.L. et al., 1992, {\it Nature}, {\bf 357}, 306

Cheng, K.S. and Ding, Winnis K.Y., 1994,  {\it Astrophys. J.}, {\bf 431}, 724

Fierro, J.M. et al., 1993, {\it Astrophys. J. Letter}, {\bf 413}, L27

Fomin, V.P. et al., 1994, {\it Astroparticle Phys.}, {\bf 2}, 137

Gregory, A.G. et al., 1990,
 {\it Proc. 21st Int. Cosmic Ray Conf. (Adelaide)}, {\bf 4}, 228

Hara, T. et al., 1993, {\it Nuc. Inst. Meth. Phys. Res. A}, {\bf 332}, 300

Kifune, T. et al., 1993a, {\it IAU Circular}, {\bf 5905}

Kifune, T. et al., 1993b,
 {\it Proc. 23rd Int. Cosmic Ray Conf. (Calgary)}, {\bf 2}, 592

Kniffen, D.A. et al., 1992, {\it IAU Circular}, {\bf 5485}

Lewis, C.W. et al., 1993, {\it Proc. 23rd Int. Cosmic Ray Conf. (Calgary)},
{\bf 1}, 279

McAdam, W.B., Osborne, J.L., and Parkinson, M.L., 1993,
 {\it Nature}, {\bf 361}, 516

Nel, H.I. et al., 1993,  {\it Astrophys. J.}, {\bf 418}, 836

Nolan, P.L. et al., 1993, {\it Astrophys. J.}, {\bf 409}, 697

Ogio, S. et al., 1993,  {\it Proc. 23rd Int. Cosmic Ray Conf. (Calgary)},
{\bf 1}, 392

Patterson, J.R. and Kifune, T., 1992, {\it Australian and New Zealand
Physicist},
{\bf 29}, 58

Punch, M. et al., 1991, {\it Nature}, {\bf 358}, 477

Swanenburg, B.N. et al., 1981, {\it Astrophys. J. Letters }, {\bf 243}, L69

Tanimori, T. et al., 1994, {\it Astrophys. J. Letters}, {\bf 429}, L61

Thompson, D.J. et al., 1992, {\it Nature}, {\bf 359}, 615

Tsukagoshi, T., 1994, {\it Thesis for Degree of Master of Science},

\hskip10truemm Tokyo Institute of Technology (in Japanese)

Vacanti, G. et al., 1991, {\it Astrophys. J.}, {\bf 377}, 467

Weekes, T.C. et al., 1989, {\it Astrophys. J.}, {\bf 342}, 379

\medskip
\vfill
\eject

\baselineskip=0.8cm

\noindent{\bf Figure legends}

\medskip

\noindent FIG.$\, 1$ \hskip1truecm The number of observed events as a function
of the
orientation angle $\alpha$ of the \v Cerenkov light image.
On- and off-source data are indicated by the solid
and dotted lines, respectively. (The on-source and off-source observing times
are not equal.)
(a) All data.
(b) The data from 1992.
(c) The data from July 1993.
(d) The data from August 1993.

\medskip

\noindent FIG.$\, 2$ \hskip1truecm  The contour map of statistical significance
for various directions in the sky around the PSR1706-44 position.
In calculating the significance,
 the assumed position
of the \lq$\gamma$-ray point source' was artificially varied
around the known position of PSR1706-44  and
the significance of any resulting peak in the $\alpha$ distribution calculated.
The result shown in the figure is from the data set of
July and August 1993.
Contours of an equal significance  are shown in the lower figures
against the two dimensional directions in units of degrees.
The upper figures are the corresponding lego plots.
In order to have the same scale both in on- and off-source plots
to facilitate comparison,
an artificial highest significance of 8.5 $\sigma$ was inserted
in the farthest left-hand bin.
The figures on the left are for on-source data and
those on the right for off-source data.

\medskip

\medskip

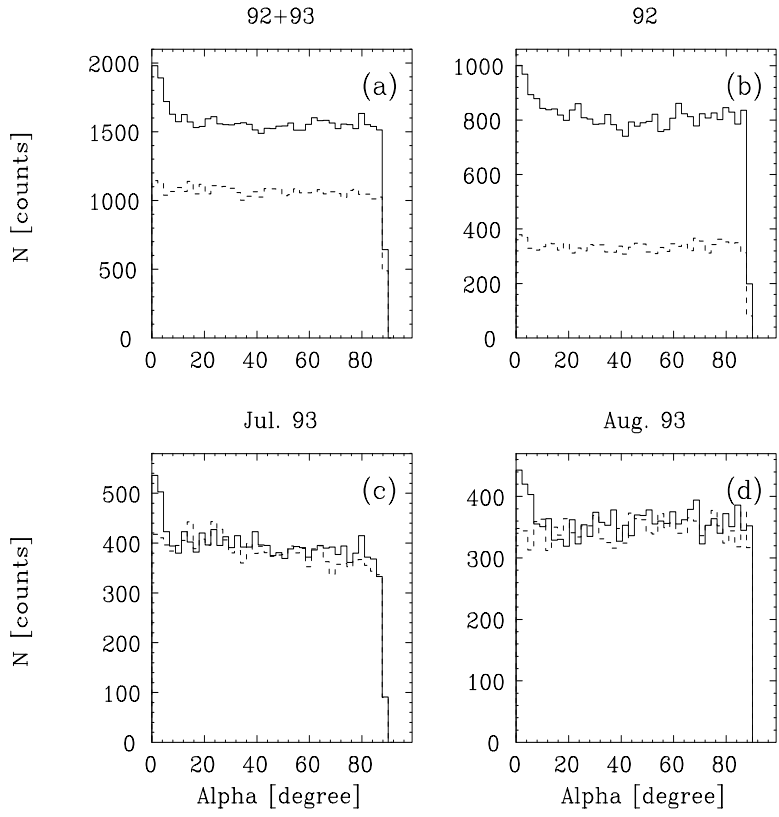
\begin{figure}[b]
\epsfysize15.5cm
\epsfbox{fig1.epsf}
\caption{}
\label{alpha}
\end{figure}

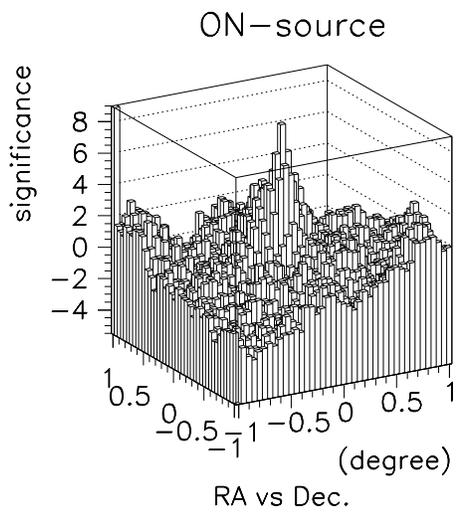
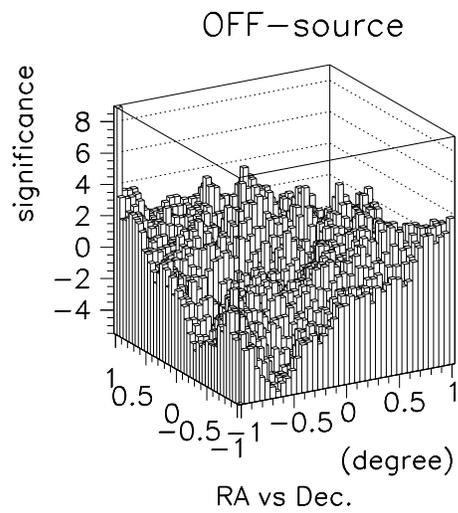
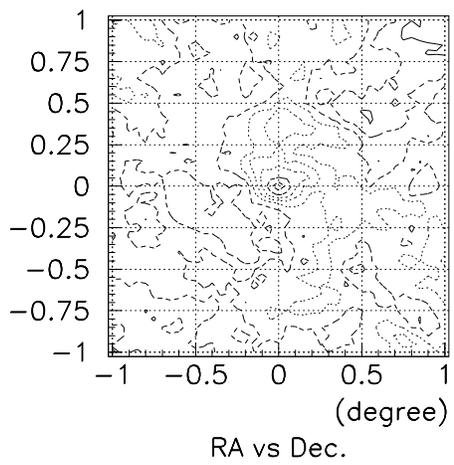
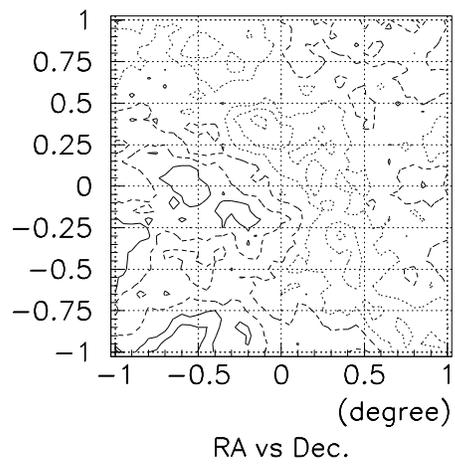
\begin{figure}[b]
\epsfysize18cm
\epsfbox{fig2.epsf}
\caption{}
\label{plane}
\end{figure}

\end{document}